\documentclass{article}

\usepackage{graphicx}
\usepackage{natbib} 
\usepackage[utf8]{inputenc} 
\usepackage[T1]{fontenc}    
\usepackage{hyperref}       
\usepackage{url}            
\usepackage{booktabs}       
\usepackage{amsfonts}       
\usepackage{amsmath}
\usepackage{amssymb}
\usepackage{mathtools}
\usepackage{amsthm}
\usepackage{mathrsfs,bm,yfonts,dsfont}
\usepackage{caption,subcaption,makecell} 
\usepackage{xcolor}
\usepackage{placeins}
\usepackage{setspace} 
\usepackage{caption} 
\newcommand{\R}{\mathbb{R}}

\newcommand{\X}{\mathbf{X}}
\newcommand{\x}{\mathbf{x}}

\newcommand{\f}{\mathbf{f}}
\newcommand{\g}{\mathbf{g}}
\newcommand{\C}{\mathbf{C}}
\newcommand{\D}{\mathbf{D}}
\newcommand{\E}{\mathbb{E}}
\newcommand{\I}{\mathbf{I}}
\newcommand{\U}{\mathbf{U}}
\newcommand{\V}{\mathbf{V}}
\newcommand{\W}{\mathbf{W}}
\newcommand{\Z}{\mathbf{Z}}
\newcommand{\Lb}{\boldsymbol{\Lambda}}
\newcommand{\w}{\mathbf{w}}

\newcommand{\dblpr}{^{\prime\prime}}

\DeclareMathOperator{\trace}{trace}
\DeclareMathOperator{\diag}{diag}

\usepackage{PRIMEarxiv}

\usepackage[utf8]{inputenc} % allow utf-8 input
\usepackage[T1]{fontenc}    % use 8-bit T1 fonts
\usepackage{hyperref}       % hyperlinks
\usepackage{url}            % simple URL typesetting
\usepackage{booktabs}       % professional-quality tables
\usepackage{amsfonts}       % blackboard math symbols
\usepackage{nicefrac}       % compact symbols for 1/2, etc.
\usepackage{microtype}      % microtypography
\usepackage{lipsum}
\usepackage{fancyhdr}       % header
\usepackage{graphicx}       % graphics
\graphicspath{{media/}}     % organize your images and other figures under media/ folder

\title{Shared Active Subspace for Multivariate
Vector-valued Functions}

\author{
  Khadija Musayeva, Micka\"el Binois \\
  Universit\'e C\^ote d’Azur, Inria, CNRS, LJAD, Sophia Antipolis, France \\
 \texttt{\{khadija.musayeva, mickael.binois\}@inria.fr} \\
}

\begin{document}
\maketitle
\begin{abstract}
This paper proposes several approaches as baselines to compute a shared active subspace for multivariate vector-valued functions. The goal is to minimize the deviation between the function evaluations on the original space and those on the reconstructed one. This is done either by manipulating the gradients or the symmetric positive (semi-)definite (SPD) matrices computed from the gradients of each component function so as to get a single structure common to all component functions.
These approaches can be applied to any data irrespective of the underlying distribution unlike the existing vector-valued approach that is constrained to a normal distribution. We test the effectiveness of these methods on five optimization problems. The experiments show that, in general, the SPD-level methods are superior to the gradient-level ones, and are close to the vector-valued approach in the case of a normal distribution. Interestingly, in most cases it suffices to take the sum of the SPD matrices to identify the best shared active subspace. 
\end{abstract}

\keywords{Dimension reduction \and symmetric positive definite matrices \and common principal component analysis \and multi-objective optimization}

\section{Introduction}

Many problems in machine learning, optimization, uncertainty quantification and sensitivity analysis suffer from the curse of dimensionality, where the performance and the complexity of the model worsens dramatically with the number of input variables. To alleviate this problem, one is interested in dimensionality reduction techniques. For instance, in machine learning, variable/feature selection methods \cite{Guyon03} aim to find a subset of variables so as to improve the predictive performance of a learning algorithm, and in some algorithms, such as decision trees, the variable selection is an inherent part of the learning process. The field of sensitivity analysis mostly deals with identifying the subset of inputs parameters whose uncertainty contributes significantly to that of the model output \cite{Saltelli08, Da21}. They are focused on the effects of the initial variables and their interactions.

However, it might be the case that the model or function of interest varies the most along directions not aligned with the coordinate axes.
The widely used dimensionality reduction method of principal component analysis (PCA) (also Karhunen-Loeve method) can be used to find a linear subspace of the input/output space containing the most of its variance, but, by default, it does not take into account the input-output relationship. In ecological sciences, the redundancy analysis applies PCA to the fitted values from a linear regression model to identify a subset of input parameters contributing significantly to the variation in the response matrix \cite{Legendre11}. This is based on the assumption of a linear relationship between the response variables and the input parameters. For regression, sufficient dimension reduction techniques aim to find a lower dimensional linear subspace such that the response and the input variables are conditionally independent given the projection of these input variables onto this subspace \cite{Cook96}. In particular, to compute such a subspace, the work of \cite{Fukumizu14} is based on the gradients of the regression function. The method of active subspace of \cite{Constantine14} also assumes that one has an access to the gradients of a function of interest, on which it performs PCA-like analysis to identify the dominant directions of variability.
But it is solely applicable to scalar-valued functions. This paper is focused on the vector-valued extension of this method.

To the best of our knowledge, only two works deal with the gradient-based dimensionality reduction for vector-valued functions. The method of \cite{Zahm20} applies only to the distributions satisfying Poincar\'e-type inequalities, primarily a normal distribution. To identify the dominant directions of variability, it takes into account both the function gradients and the data covariance matrix. On the other hand, \cite{Ji18} extends the active subspace method rather trivially: it is based on a linear combination of the active subspaces computed separately for each output using the method of \cite{Constantine14}, and applies only to a normally distributed data as well. In this paper, irrespective of the data generating distribution, we aim to reduce the vector-valued problem to the scalar setting either at the level of gradients, or at the level of symmetric positive definite (SPD) matrices (i.e., covariance-like matrices), computed from these gradients.
The former group of methods aims to compute a single representative vector for the component functions' derivatives with respect to each input variable, such as taking their average or projecting them to a one dimensional linear subspace. The SPD-level methods range from taking the sum of SPD matrices, to tools from the common principal component analysis (\cite{Flury88},\cite{Trendafilov21}) to find a common structure in multiple SPD matrices. Particularly, the method of \cite{Flury86} deals with the joint diagonalization of multiple SPD matrices and that of \cite{Trendafilov10} with stepwise estimation of common structures in those matrices. The eigenvectors obtained from these methods then represents the important directions of variability. The experiments on one synthetic and four real-world inspired optimization problems show that the latter group of methods, in particular, the sum of SPD matrices, are  very effective approaches to computing a common active subspace for vector-valued functions with respect to the root-mean-square error. In particular, for a normally distributed data their performances are very close to that of the method of \cite{Zahm20}.

The paper is organized as follows. Section~\ref{sec:methodology} gives a general overview of the active subspace method of \cite{Constantine14}, and that of \cite{Zahm20} for vector-valued functions. In Section~\ref{sec:our_approach}, we introduce and discuss our approaches to compute a shared active subspace for vector-valued functions. Section~\ref{sec:exp} discusses the numerical experiments conducted to test their effectiveness, and the results obtained.
Section~\ref{sec:conclusions} concludes the paper with a brief discussion of possible research directions on this topic. We provide an \texttt{R} package implementation of the proposed methods as well as that of \cite{Zahm20} in the supplementary material.

\section{Methodological background}
\label{sec:methodology}
In what follows, we denote by $\mathcal{X} \subseteq \R^d$ an input/parameter space and $\mathbb{P}$ is a probability distribution with density $\rho$ defined on $\left(\mathcal{X}, \mathcal{B}(\mathcal{X})\right)$, where $\mathcal{B}(\mathcal{X})$ are Borel sets of $\mathcal{X}$. $\X$ stands for a random variable with the values in $\mathcal{X}$ and $\D_n =\{\x_1, \dots, \x_n\} \subset \mathcal{X}$ is an $n$-sample distributed according to $\mathbb{P}^n$. Let $\mathcal{H}$ denote the space of square-integrable  functions $\f=(f_1, \dots, f_C) : \mathcal{X} \rightarrow  \R^C$ with respect to $\rho$, where $f$ stands for a real-valued function, with the norm $\lVert \f \rVert^2_{\mathcal{H}} = \E_{\X \sim \mathbb{P}}[\f(\X) \f(\X)^T]$. We denote the gradient of the $k$-th component of $\f$, $k \in \{1, \dots, C\}$, by $\displaystyle{\nabla f_k(\x) = \left[
\frac{\partial f_k(\x)}{\partial \x^{(1)}} \dots\frac{\partial f_k(\x)}{\partial \x^{(d)}}\right]}$, where $\x^{(i)}$ is the $i$-th element of the input vector $\x$, and we denote the gradient evaluations on $\D_n$ by $\displaystyle{\nabla_{\D_n} f_k = \left[(\nabla f_k(\x_1))^T \dots (\nabla f_k(\x_n))^T\right]^T}$. $\mathbf{J}$ stands for the Jacobian of $\f$, i.e.,
$
\mathbf{J} = \begin{bmatrix} 
\frac{\partial f_1(\x)}{\partial \x^{(1)}} & \dots & \frac{\partial f_1(\x)}{\partial \x^{(d)}} \\
\vdots & \ddots & \vdots \\
\frac{\partial f_C(\x)}{\partial \x^{(1)}} & \dots & \frac{\partial f_C(\x)}{\partial \x^{(d)}}
\end{bmatrix}.
$ With this notation at hand, the next two sections give a brief overview of the existing active subspace methods.

\subsection{Active subspace} \label{section:single-objective}

The goal of the active subspace method of \cite{Constantine14}, on which our work is based, is to find the directions of the strongest variability of $f$ based on the eigenvalue decomposition of the SPD matrix (which can be thought of as an uncentered covariance matrix),  $\C = \E_{\X \sim \mathbb{P}}[(\nabla f(\X))^T \nabla f(\X)]$. In the obtained decomposition, $\C = \W \Lb \W^T$, provided that there is a sharp decay in the eigenvalues in the form $\Lb=\diag(\Lb_1,\Lb_2)$, where $\Lb_1=[\lambda_{1} \dots \lambda_l]$ and $\Lb_2=[\lambda_{l+1} \dots \lambda_d]$, we can decompose $\W = \left[ \W_1 \; \W_2\right]$. The eigenvectors $\W_1$ correspond to the largest eigenvalues $\Lb_1$ and identify the directions of the strongest average variability of $f$, i.e., $f$ varies stronger on the coordinates $\x' = \W_1^T \x$ than on the $\x\dblpr = \W_2^T \x$. Then, for a fixed $\x'$, $f(\x)=f(\W_1 \x' + \W_2 \x\dblpr)$ can be approximated by averaging out over all values of $\x$ that map to $\x'$ as $\hat{f}(\x') = \E_{\X\dblpr \sim \mathbb{P}^{'}_{\X\dblpr | \X'}}\left[f\left(\W_1 \x' + \W_2 \X \dblpr \right)\right]$, where $\mathbb{P}^{'}$ is the joint distribution of the pair $(\X', \X\dblpr)$ with the density $\rho(\W_1 \x' + \W_2 \x \dblpr)$.

Since the computation of $\C$ and $\hat{f}$ involve that of high-dimensional integrals, in practice they are estimated based on data, or on a surrogate model, see e.g., \cite{wycoff2021sequential}. First, $\W$ is estimated from the eigendecomposition of
$\displaystyle{\hat{\C}= \frac{1}{n} \sum_{i=1}^n (\nabla f(\x_i))^T \nabla f(\x_i)}$ computed using $\D_n$. This is, in fact, similar to performing PCA on the gradient evaluations $\nabla_{\D_n} f$. The obtained eigenvectors $\hat{\W}$ are then used to approximate $f$ as 
\begin{align}
\hat{f}(\x') = \frac{1}{N}\sum_{i=1}^N\left[f\left(\hat{\W}_1 \x' + \hat{\W}_2 \x \dblpr_i \right)\right], \label{eq:single-objective}
\end{align}
where $\x\dblpr_i$ is sampled independently from the marginal $\tilde{\mathbb{P}}'_{\X\dblpr | \X'}$ of $\tilde{\mathbb{P}}'$ with the density $\rho(\hat{\W}_1 \x' + \hat{\W}_2 \x\dblpr)$. Under the assumption that there is an $\epsilon>0$ such that $\lVert \W- \hat{\W}\rVert \leqslant \epsilon$ in the matrix $2$-norm, for this (conditional expectation) approximation of $f$, using the Poincar\'e-type inequalities the authors derive an upper bound on the mean squared error as  
$$ \E_{\X \sim \mathbb{P}}\left[ (f(\X)-\hat{f}(\hat{\W}^T_1\X))^2\right] \leqslant K \left(1+\frac{1}{N}\right)\left(\epsilon \left(\sum_{i=1}^{l} \lambda_i\right)^{1/2} + \left(\sum_{j=l+1}^{d}\lambda_j\right)^{1/2}\right)^2,$$
where $K$ is a constant depending on $\mathcal{X}$ and $\mathbb{P}$. Given that the dependence on $N$ is negligible, if $\epsilon$ is small and if $[\lambda_{l+1} \dots \lambda_d]$ decays rapidly, this bound will be small. This motivates us to consider the approximation of $f(\x)$ as $f(\hat{\W}_1 \x') = f(\hat{\W}_1 \hat{\W}_1^T \x)$ using only the dominant eigenvectors $\hat{\W}_1$ of $\hat{\W}$. The question is then the sample size $n$ needed to obtain an $\epsilon$-approximation of $\W$. To this end, \cite{Holodnak18} provides a lower bound on the sample size $n$ in terms of $\epsilon$ and the quantities characterizing the smoothness of $f$. Assuming that $f$ is smooth enough, this dependence takes the form $\mathcal{O}(1/\epsilon^2)$.

It should be noted that \cite{Lee19} proposed a modification to this active subspace computation by taking into account the average of gradients $\Z=\E_{\X \sim \mathbb{P}}[(\nabla f(\X))]$ as $\C = \E_{\X \sim \mathbb{P}}[(\nabla f(\X))^T \nabla f(\X)]+\Z^T\Z$. This is shown to be an efficient dimensionality reduction approach for quadratic functions with linear trends from the perspective of optimizing the (un)explained variance of $f$.

\subsection{Vector-valued active subspace} \label{section:multi-objective}

Under the assumption that $\mathbb{P}$ is a Gaussian distribution $\mathcal{N}(\mu, \Sigma)$ (or any other distribution that satifies Poincar\'e-type inequalities), \cite{Zahm20} propose to identify the directions of variability by approximating $\f$ by a ridge function of the form $\g \circ \mathbf{P_r}$, where $\mathbf{P_r} \in \R^{d \times d}$ is a projector of rank $r$ (ideally $r \ll d$), and $\g : \R^d \rightarrow \R^C$, so that 
$\displaystyle{ \lVert \f - \g \circ \mathbf{P_r} \rVert_{\mathcal{H}} }$ is minimum. For any $\x \in \mathcal{X}$, such a minimizer takes the form of $\E_{\X^{\prime} \sim \mathbb{P}} \left[ \f(\mathbf{P_r} \x+ (\I-\mathbf{P_r})\X^{\prime}) \right]$, i.e., the orthogonal projection of $\f$ onto the set $\left\{\g \circ \mathbf{P_r} \mid \g : \R^d \rightarrow \R^C \right\} \cap \mathcal{H}$, where $\I$ is the identity matrix of dimension $d$. Under the assumption that $\f$ is continuously differentiable, similarly to the scalar-valued case, the upper bound on such an approximation of $\f$ can be obtained thanks to the Poincar\'e-type inequalities as  \begin{align} 
\lVert \f -  \E_{\X^{\prime} \sim \mathbb{P}} \left[ \f(\mathbf{P_r} \x+ (\I-\mathbf{P_r})\X^{\prime}) \right]\rVert^2_{\mathcal{H}} \leqslant \trace(\Sigma (\I-\mathbf{P_r}^T) \mathbf{H} (\I-\mathbf{P_r})),\label{eq:zahm_bound}
\end{align}
where $\mathbf{\Sigma}$ is a (non-singular) covariance matrix and $\mathbf{H}$ can be expressed as $\E_{\X \sim \mathbb{P}}\left[ \mathbf{J}^T(\X) \mathbf{J}(\X)\right]$. The projector $\mathbf{P_r}$ minimizing \eqref{eq:zahm_bound} is then derived in terms of the generalized eigenvectors $\mathbf{v}_i \in \R^d$ of the pair $(\mathbf{H}, \mathbf{\Sigma}^{-1})$ as $\mathbf{P_r}=\left(\sum_{i=1}^r \mathbf{v}_i \mathbf{v}_i^T\right)\mathbf{\Sigma}^{-1}$. Thus, the bound \eqref{eq:zahm_bound} is characterized not only by the gradients of $\f$, but also by the covariance structure of $\X \sim \mathbb{P}$. Similarly to the single-objective case, since \ref{eq:zahm_bound} involves high-dimensional integrals, $\f$ is approximated based on data as follows. First, $\mathbf{H}$ is estimated as $\displaystyle{\hat{\mathbf{H}} = \frac{1}{n} \sum_{i=1}^n \mathbf{J}^T(\x_i) \mathbf{J}(\x_i)}$  based on $\D_n$. The projection $\mathbf{\hat{P}}_r$ is then computed based on the generalized eigendecomposition of the pair $(\hat{\mathbf{H}}, \mathbf{\Sigma}^{-1})$. Finally, for any $\x$, $\E_{\X^{\prime} \sim \mathbb{P}} \left[ \f(\mathbf{P_r} \x+ (\I-\mathbf{P_r})\X^{\prime})\right]$ is estimated based on $N$-sample $(\x^{\prime}_1, \dots, \x^{\prime}_N)$ distributed according to $\mathbb{P}^N$ as
\begin{align}
\hat{\f}(\x)=\frac{1}{N} \sum_{i=1}^N \f(\mathbf{\hat{P}}_r \x+ (\I-\mathbf{\hat{P}}_r)\x^{\prime}_i).
\label{eq:zahm_bound_data_based}
\end{align}
One can notice a similar form to the single-objective case approximation \eqref{eq:single-objective}, where now the second part being an orthogonal projection does not depend on the first part, thus the same $N$-sample is used for any $\x$.

Another active subspace approach for vector-valued functions is that of \cite{Ji18}. Under the assumption that $\mathbb{P}$ is the standard Gaussian distribution $\mathcal{N}(\mathbf{0}, \I)$, their goal is to estimate the marginal distributions of $f_k(\X)$, $1 \leqslant k \leqslant C$, based on the active subspace approach. This work consists of, first, computing the leading eigenvectors $\W_k$ for each $f_k$ separately based on the work of \cite{Constantine14} as discussed in Section~\ref{section:single-objective}, then finding the matrix $\V \in \R^{d \times r}$, where $r$ is the dimension of the active subspace assumed to be the same for all $f_k$, so that $\displaystyle{\W^T_k \V}=\I$, i.e., $\V$ is common to all component functions. Such a matrix gives the identity $f_k(\x)=f_k(\V \W^T_k \x)$. Then, under the assumption that $\X \sim \mathcal{N}(\mathbf{0}, \I)$,  $\W^T_k \X$ also follows the standard Gaussian distribution and thus $f_k(\X)$ is distributed in the same way as $f_k(\V \W^T_k \X)$.

%For a linear function $\f(\x) = \mathbf{A} \x $ with $\mathbf{A} \in \R^{d \times d}$, \eqref{eq:zahm_bound} holds with equality.

In the next section, we propose new approaches to shared active subspace computation for vector-valued functions which do not depend on the data generating distribution.

\section{New baselines to vector-valued active subspace}
\label{sec:our_approach}

The straightforward vector-valued extension of \cite{Constantine14} consists of treating each component of $\f=(f_1, \dots, f_C)$ as a separate problem, and computing an active subspace for each separately. But this is limiting (unless some functions share exactly the same active directions, which is a rather restrictive assumption), since any joint analysis would need to be performed on the concatenation of all active subspaces. 

Our goal here is to find eigenvectors $\hat{\W}_j$, $1 \leqslant j \leqslant d$, common to all objectives $f_1, \dots, f_C$, irrespective of the distribution $\mathbb{P}$, so as to minimize 
\begin{align}
\sum_{k=1}^C \left(\frac{1}{n}\sum_{i=1}^n (f_k(\x_i)-f_k(\hat{\W}_j \hat{\W}_j^T \x_i))^2\right)^{1/2}. \label{err:sum_root_mean_square_error}
\end{align}
We do this by 
transforming the vector-valued problem into the scalar-valued one of Section \ref{section:single-objective} either at the level of the gradients, where the Jacobian at each point of $\D_n$ is replaced by a $d$-dimensional vector, or at the level of the SPD matrices by manipulating the component-wise SPD matrices $\displaystyle{\hat{\C}_k= \frac{1}{n} \sum_{i=1}^n (\nabla f_k(\x_i))^T \nabla f_k(\x_i)}$, $1 \leqslant k \leqslant C$. In a sense, these approaches are in the spirit of the work of \cite{Ji18},  where the combination is now performed at the level of gradients $\nabla f_k$ or SPD matrices $\hat{\C}_k$, $1 \leqslant k \leqslant C$, instead of eigenspaces.

\subsection{Manipulation at the level of gradients} \label{sec:manip_gradients}

 In this section, we propose several approaches to reduce the Jacobian evaluated at each point $\x$ in $\D_n$ to a $d$-dimensional vector $\mathbf{u}(\x)$. Once such a vector is computed, one can proceed as in Section \ref{section:single-objective}.

\subsubsection{Linear projection of gradients}
One of the ways of doing so is to project the derivatives of each objective with respect to (wrt) each input parameter $\x^{(j)}$, $j \in \{1, \dots, d\}$, to the one-dimensional subspace that captures the most of the variance in these derivatives, i.e., to perform PCA-like analysis. In detail, we combine the derivative evaluations on $\D_n$ of each component function $f_k$, $1 \leqslant k \leqslant C$, wrt the $j$-th parameter into a single matrix
$\displaystyle{
\mathbf{J}_j = \begin{bmatrix} 
\frac{\partial f_1(\x_1)}{\partial \x^{(j)}_1} & \dots & \frac{\partial f_C(\x_1)}{\partial \x^{(j)}_1} \\
\vdots & \ddots & \vdots \\
\frac{\partial f_1(\x_1)}{\partial \x^{(j)}_n} & \dots & \frac{\partial f_C(\x_n)}{\partial \x^{(j)}_n}
\end{bmatrix}
}$. Then, for each $\mathbf{J}_j$, we find a direction $\w_j \in \R^{C \times 1}$ capturing the most of the variability in $\mathbf{J}_j$ from the eigendecomposition of the SPD matrix $\displaystyle{\frac{\mathbf{J}^T_j \mathbf{J}_j}{n} \in \R^{C \times C}}$ and approximate $\mathbf{J}_j$ as $\mathbf{J}_j \w_j$. 
Then, the $i$-th row in the matrix $\mathbf{A} = \left[\mathbf{J}_1 \w_1 \dots \mathbf{J}_d \w_d \right] \in \R^{n \times d}$ combining these rank-1 projections can be seen as a representative vector for the Jacobian evaluated at $\x_i$. The eigendecomposition of $\mathbf{A}^T\mathbf{A}/n$ yields the eigenvectors common to all objectives. For a given input variable, this approach will give the most of the weight to the component function that highly varies wrt this variable.  

\subsubsection{Convex hull of gradients}
We notice that the problem we are interested in bears similarity to finding common descent direction for multiple objectives in the multi-objective optimization task. \cite{Desideri12} shows that 
if there is no zero convex combination of gradient vectors $\nabla f_k(\x)$, i.e., if there does not exist a positive $\alpha \in \R^C_{+}$ with $\alpha\mathbf{\alpha}^T = 1$ satisfying
$\alpha \mathbf{J} = \mathbf{0}$, then a descent direction from this point common to all gradient vectors exists. It is calculated based on the vector $\mathbf{u}(\x) \in \R^d$ satisfying $\mathbf{J} \mathbf{u}(\x)^T \geqslant \mathbf{0}$. Furthermore, 
such a vector can be found by computing the minimum (Euclidean) norm element of the convex hull of the set $\{\nabla f_1(\x), \dots, \nabla f_C(\x)\}$,
$$\U = \left\{\mathbf{u}(\x) \in \R^d \mid \mathbf{u}(\x) = \sum_{k=1}^C \alpha_k \nabla f_k(\x), \forall k, \alpha_k \geqslant 0, \sum_{k=1}^C \alpha_k = 1 \right\}.$$ 
Here, we are not interested in the existence of the common descent direction since we are not performing optimization, but only in finding the vector $\mathbf{u}(\x)$ at each point $\x$. %We can find the minimum norm element of $\U$ using the algorithm of \cite{Sekitani93}.
%where $\nabla f_k(\x) = \nabla f_k(\x)/\lVert \nabla f_k(\x) \rVert_2$ is the normalized vector.
A more egalitarian approach would be to simply consider the average vector $$\mathbf{u}(\x) =\frac{1}{C} \sum_{k=1}^C \nabla f_k(\x),$$ which is also an element of $\U$. However, it is not necessarily positively correlated with the gradient vectors, as it is the case for the minimum norm element of $\U$. On the other hand, if $\mathbf{u}(\x)$ is the average of the normalized gradients, $\mathbf{J}'_{k.}=\nabla f_k(\x)/\lVert \nabla f_k(\x) \rVert_2$, it will satisfy $\mathbf{J}' \mathbf{u}(\x)^T \geqslant \mathbf{0}$, where $\mathbf{J}'_{k.}$ stands for the $k$-th row of $\mathbf{J}'$. However, normalizing the gradients does not suit the context of this work, because it changes their magnitudes which determine the active subspace. Given an input variable, if the linear projection of the gradients favors the highly varying objective and the minimum norm element of the convex hull the least varying one, the average will provide a compromise solution between these two.

In this approach, a shared eigenspace is computed from the eigendecomposition of  $\displaystyle{\frac{1}{n} \sum_{i=1}^n \mathbf{u}(\x_i)^T\mathbf{u}(\x_i)}$.
%is a normalized Jacobian with the $k$-th row equal to $\mathbf{J}'_{k.} = \nabla f_k(\x)/\lVert \nabla f_k(\x) \rVert_2$, will be positively correlated with all components $\nabla f_k(\x)$. 
%Such a vector can be easily found by simply taking the average $ \displaystyle{\mathbf{u}(\x) =\frac{1}{C} \sum_{k=1}^C \mathbf{J}'_{k.}}$. 
%In each case we compute $\hat{\C}=\mathbf{A}^T \mathbf{A}$ and proceed as in the single-objective case. 

\subsection{Manipulation at the level SPD matrices} \label{sec:manip_SPD}
We can compute a common eigenspace by combining the component matrices  $\hat{\C}_k$, $1 \leqslant k \leqslant C$, into a single SPD matrix $\hat{\C}$ and eigendecomposing it. A more involved approach is related to the common principal component analysis, a generalization of PCA to several covariance matrices where the goal is to find common eigenvectors to all of them. %Using this method we can find such vectors for the SPD matrices $\hat{\C}_k$.

\subsubsection{Sum of SPD matrices}
The straightforward approach to obtain a single common SPD matrices is to consider the sum $\hat{\C}=\sum_{k=1}^C \hat{\C}_k$, which is itself an SPD matrix.
Notice that, for the standard normal distribution, this method will provide similar results to that in \eqref{eq:zahm_bound_data_based}, since in this case, the projector becomes
$P_r=\sum_{i=1}^r \mathbf{v}_i \mathbf{v}_i^T$ which is the same as the sum just mentioned. In fact, \cite{Krzanowski79} has communicated the effectiveness of the average of the sample covariance matrices in the estimation of approximate common principal components. It should be noted that any scaling $\alpha \hat
{\C}$, with $\alpha>0$, will only scale the eigenvalues as $\Lb/\alpha$ leaving intact the eigenvectors, consequently, it has no effect on the computation of the dominating eigenspace.

\subsubsection{Joint diagonalization of SPD matrices}
A more involved approach to find a common structure in $\{\hat{\C}_1, \dots, \hat{\C}_C\}$ is based on the joint diagonalization of these matrices \cite{Flury86,Flury87}. This method has been used for estimating the common principal components for multiple groups and a series of works has been dedicated to this line of research (\cite{Beaghen97,Trendafilov10}, \cite{Browne14}, \cite{Trendafilov21} and references therein). If all $\hat{\C}_k$ are simultaneously diagonalizable, the equality $\W^T \hat{\C}_k \W = \Lb_k$, where $\Lb_k$ is diagonal, for all $k$ holds. But usually this is not the case and \cite{Flury86} (FG) proposed an iterative algorithm to compute $\hat{\W}$ such that the matrices $\hat{\W}^T \hat{\C}_k \hat{\W}$ are as diagonal as possible by minimizing the deviation from the diagonality,
\begin{align}
\prod_{k=1}^C \left(\det(\diag(\hat{\W}^T \hat{\C}_k \hat{\W}))/\det(\hat{\W}^T \hat{\C}_k \hat{\W})\right)^{n_k}, \label{problem:FG}
\end{align}
where $\det(\cdot)$ stands for the determinant and $n_k$ is the number of samples in the group $k$, which in our case is equal to $n$. The quantity \eqref{problem:FG} appears as a test statistic of the simultaneous diagonalization of the covariance matrices of multiple normally distributed random variables \cite{Flury1984}. 

The specifics of the FG algorithm is that although it finds a common structure in $\hat{\C}_k$'s, it does not produce any canonical ordering on $\hat{\W}$. In other words, although the eigenvectors are common to all components, the decay structure is not necessarily the same for all $\Lb_k$. Hence, to identify the active subspace, we need to induce an ordering on the columns of $\hat{\W}$. 
Although different weighting approaches can be applied to induce an ordering on the eigenvalues from the FG algorithm, to keep it simple, in this paper we identify the dominating subspace based on the ordering induced by the sum $\sum_{k=1}^C \Lb_k$.

\subsubsection{Stepwise estimation of eigenvectors}
In fact, due to the arbitrariness of the ordering of eigenvectors, the FG method has been considered unsuitable for the dimensionality reduction task. Instead, \cite{Trendafilov10} proposed an approach to find approximate common eigenvectors in a stepwise manner so that all groups share approximately the same decay structure. He shows that the minimization problem \eqref{problem:FG} is equivalent to 
$$\sum_{k=1}^C n_k \log(\det(\diag(\hat{\W}^T \hat{\C}_k \hat{\W}))),$$
which can be expressed in the vectorized form as
$$\sum_{k=1}^C n_k \sum_{j=1}^d\log(\hat{\w}^T_j \hat{\C}_k \hat{\w}_j).$$ The latter quantity can then be minimized by sequentially minimizing the $d$ identical problems
of the form $$\sum_{k=1}^C n_k \log(\hat{\w}_j^T \hat{\C}_k \hat{\w}_j), \; 1 \leqslant j \leqslant d.$$
More precisely, each $\hat{\w}_j$ is found in the orthogonal subspace of the previous solution $\hat{\w}_{j-1}$, thus yielding an ordering on the decreasing importance of these vectors, and this ordering holds simultaneously for all $\{\hat{\C}_1, \dots, \hat{\C}_C\}$ . The approach of Trendafilov yields the same result as that of FG method, if the eigenvalues $\Lb_k$ decreases simultaneously for all $k$ groups.

%We would like to find the ordering so that the eigenspace $\hat{\W}_1$ chosen according to this ordering minimizes the error: $$\displaystyle{\frac{1}{n}\sum_{i=1}^n \sum_{k=1}^C (f_k(\x_i)-f_k(\hat{\W}_1 \hat{\W}_1^T \x_i))^2}.$$

\section{Numerical Experiments}
\label{sec:exp}

Our experiments include one synthetic, and the following four real-world inspired optimization problems: automobile car side-impact problem \cite{Deb09}, conceptual marine design problem \cite{Parsons04,Tanabe20}, penicillin production \cite{Liang21} and switching ripple suppressor design problem \cite{Zhang19,He20}. In the following sections, we detail the experimental setup and give a general overview of the problems  considered.

\subsection{Setup}

We performed two sets of experiments with different data distributions. In the first set, to compare the results with those of the method of \cite{Zahm20}, the data were sampled from the standard normal distribution. In the second set, since the input variables in the test problems are independent and all values are equally worth, we use the uniform distribution on $[-1, 1]^d$. In each problem, we have $\mathcal{X} = [u_1, l_1] \times \dots \times [u_d, l_d]$, and this requires an additional step to map the data to the hypercube $[0, 1]^d$ and apply the transformation $x^{(i)}\cdot(u_i - l_i) + l_i$ for all $i$. To perform this mapping, we used the sigmoid function for the normal distribution, and $(x^{(i)}+1)/2$ for all $i$ for the uniform one. We also mean-variance normalized the function outputs. 

Active subspace is based on the gradient computations which can be a costly task with a large number of input variables. Thus to have a good estimation of the eigenspace of dimension $d'$, \cite{Constantine14b,Constantine15} 
propose to use $\alpha d' \log(d)$ number of samples, where $\alpha$ varies between $2$ and $10$. Since in our case for all problems $d<10$, using samples of size $1000$ we could compute the gradients in a matter of seconds using an ordinary laptop. We repeated the experiments $10$ times. In the method of \cite{Zahm20}, for all problems, the number $N$ of samples in Equation \eqref{eq:zahm_bound_data_based} was set to $20$ (the maximum value used in the original paper).

We provide the implementation of the 
proposed approaches as an \texttt{R} package \texttt{SharedAS}. We compute the gradients using the \texttt{R} package \texttt{numDeriv}\cite{Gilbert09}. For the FG algorithm and the stepwise estimation of the eigenvectors we use the \texttt{R} package \texttt{cpc} \cite{Pepler14}. The FG method, by default, returns the eigenvectors in the order of the increasing sum of the component eigenvalues. It can be estimated either by maximum likelihood (the original approach) or least squares proposed by \cite{Beaghen97} which we actually used. We computed the minimum element of the convex hull based on the algorithm of \cite{Sekitani93}. 

\subsection{Test problems}
The problems we considered are listed in Table~\ref{table:test-problems}.

\begin{table}
\centering
\footnotesize
\begin{tabular}{ccc}
Problem & number of input variables & number of objectives \\
\hline
Synthetic problem  &  $3$ & $2$\\
Automobile car side-impact & $7$ & $11$ \\
Conceptual marine design & $6$ & $3$\\
Penicillin production & $7$ & $5$\\
Switching ripple suppressor design & $8$ & $5$ and $10$\\
\end{tabular}
\caption{Summary of the test problems.}
\label{table:test-problems}
\end{table}

\paragraph{Synthetic problem} As a synthetic problem we consider a modification of the multi-objective test problem defined by \cite{parr2013improvement}: let $b_1 = 15 x_1 - 5$, $b_2 = 15x_2$ and $b_3 = x_3$,
$$f_1(\mathbf{x}) = \left(b_2 - 5.1 (b_1/(2 \pi))^2 + 5/\pi  b_1 - 6\right)^2 + 10  \left((1 - 1/(8 \pi)) \cos(b_1) + 1\right) + \sin(\pi b_3)$$ 
and
\begin{align*}
\begin{split}
f_2(\mathbf{x}) = -\sqrt{(10.5 - b_1)  (b_1 + 5.5)  (b_2 + 0.5)} - \frac{1}{30} \left(b_2 - \frac{5.1}{4\pi^2} b_1^2 - 6\right)^2 \\ - \frac{1}{3}  \left[\left(1 - \frac{1}{8 \pi}\right) \cos(b_1) + 1\right] -
    \cos(2 \pi b_3).
\end{split}
\end{align*}
Each of these 3-dimensional functions is rotated with a different rotation matrix: the first one is 
$\mbox{\footnotesize$
  \begin{pmatrix}
    -0.71& 0.34& 0.62\\
    0.28 & 0.94 & -0.2 \\
    -0.65 & 0.032 & -0.76
\end{pmatrix}$}$, and the second one
$\mbox{\footnotesize$
 \begin{pmatrix}
    -0.32& 0.84 & -0.44\\
    0.76 & -0.058 & -0.65 \\
    0.57 & 0.54 & 0.62
\end{pmatrix}$}$.

%{\em should this be discussed in more detail?} %[\emph{BTW, can we use these matrices to define a ground truth?}]

\paragraph{Automobile car side-impact problem \cite{He20}} This problem of vehicle performance optimization involves 7 decision variables for thicknesses of car body parts (4 additional noise variables are kept fixed) and 11 objectives for the car mass, deceleration and structural integrity.

\paragraph{Conceptual marine design problem \cite{Tanabe20}} This is a simplified commercial ship sizing problem with $6$ input parameters, and $3$ objectives which are the transportation cost, light ship weight and annual cargo capacity.

\paragraph{Penicillin production problem \cite{Liang21}} is based on $7$ input variables, which include culture medium volume $V$, biomass concentration $X$, %temperature $T$, 
glucose substrate concentration $S$, %substrate feed rate $F$, substrate feed concentration $s_{f}$ and $\mathrm{H}^{+}$ concentration $H$ 
and three objectives which are penicillin concentration $P$, carbon dioxide concentration $C O_{2}$ and the reaction time $t$. Since the latter is a discrete quantity, we exclude it from our computation. In our experiments, everything else being equal to that of the original setting, we set the reaction time $t=100$, and remove all the outputs where the loop over $t$ terminates due to the extreme values of $S$ and $V$. Since $V$, $X$ and $S$ are differentiable quantities we used them as outputs, along with $P$ and $C O_{2}$, thus dealing with a problem with $5$ objectives. For this problem, we used the implementation provided on github.com/HarryQL/TuRBO-Penicillin.

\paragraph{Switching ripple suppressor design \cite{Zhang19}} The problem of the switching ripple
suppressor design for voltage source inversion in powered system is represented by $8$ input variables and $5$ objectives. The input variables are the design parameters of the electronic components,
%: inverter-side inductor $L1$, splitting grid-side inductor $L2$, grid-side inductor $L3$,switching ripple suppressor capacitor $C_f$, and resonant capacitor $C_1, \dots, C_4$.
and the objectives are the total cost of the inductors and the harmonics attenuations at two different resonant frequencies. As it is the case with the penicillin production problem, here we also used some of the constraints as outputs, thus increasing the number of the objectives to $10$. Overall, for this problem, we experimented separately with $5$ and $10$ objectives.

\subsection{Results}

\begin{figure}[htbp]
\centering
\begin{subfigure}{.5\textwidth}
\caption{Synthetic problem}
\includegraphics[width=\textwidth]{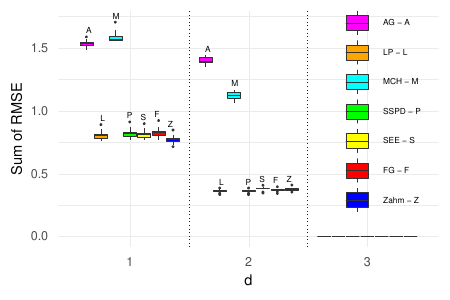}
\end{subfigure}%
\begin{subfigure}{.5\textwidth}
\caption{Car side-impact problem}
\includegraphics[width=\textwidth]{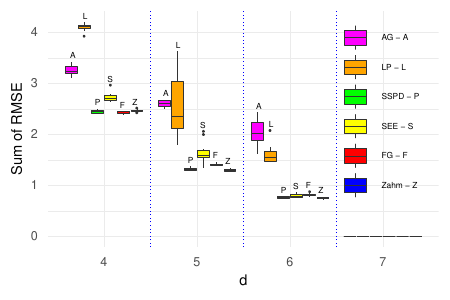}
\end{subfigure}
\begin{subfigure}{.5\textwidth}
\caption{Conceptual marine design problem}
\includegraphics[width=\textwidth]{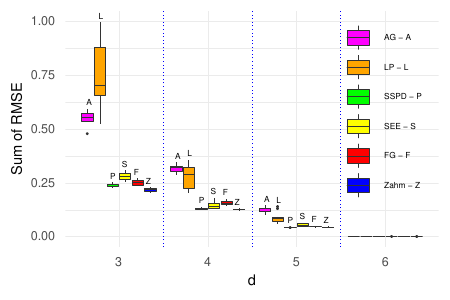}
\end{subfigure}%
%\begin{subfigure}{.5\textwidth}
%\caption{Penicillin production problem: $2$ objectives}
%\includegraphics[width=\textwidth]{fig_pen2obj_norm.pdf}
%\end{subfigure}
\begin{subfigure}{.5\textwidth}
\caption{Penicillin production problem}
\includegraphics[width=\textwidth]{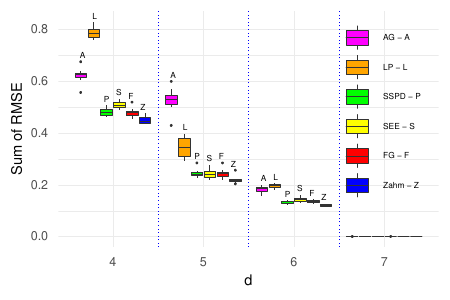}
\end{subfigure}
\begin{subfigure}{.5\textwidth}
\caption{Switching ripple: 5 objectives}
\includegraphics[width=\textwidth]{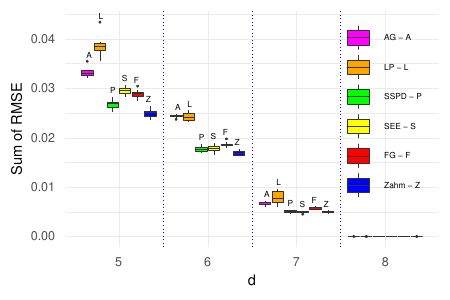}
\end{subfigure}%
\begin{subfigure}{.5\textwidth}
\caption{Switching ripple: 10 objectives}
\includegraphics[width=\textwidth]{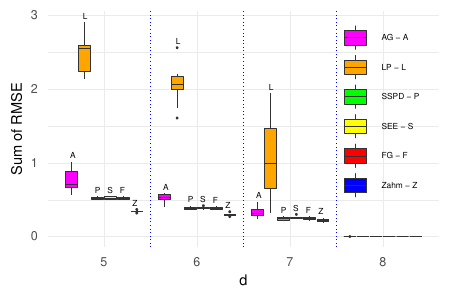}
\end{subfigure}
\caption{Comparison of the proposed approaches and the method of Zahm based on data sampled from the standard normal distribution. The experiments were repeated $10$ times with the sample size $n=1000$ in each of them.}
\label{fig:boxplot_as_normal}
\end{figure}

\begin{figure}[htbp]
\centering
\begin{subfigure}{.5\textwidth}
\caption{Synthetic problem}
\includegraphics[width=\textwidth]{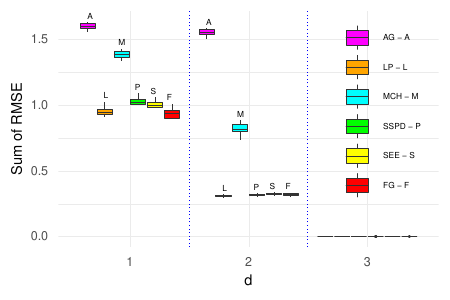}
\end{subfigure}%
\begin{subfigure}{.5\textwidth}
\caption{Car side-impact problem}
\includegraphics[width=\textwidth]{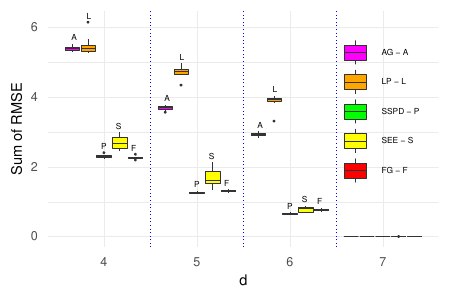}
\end{subfigure}
\begin{subfigure}{.5\textwidth}
\caption{Conceptual marine design problem}
\includegraphics[width=\textwidth]{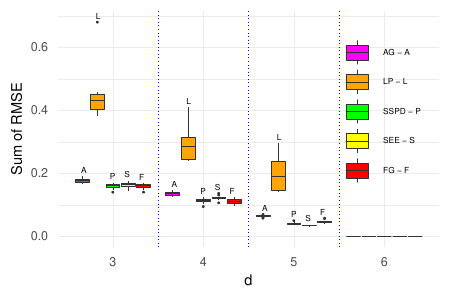}
\end{subfigure}%
%\begin{subfigure}{.5\textwidth}
%\caption{Penicillin production problem: $2$ objectives}
%\includegraphics[width=\textwidth]{fig_pen2obj_norm.pdf}
%\end{subfigure}
\begin{subfigure}{.5\textwidth}
\caption{Penicillin production problem}
\includegraphics[width=\textwidth]{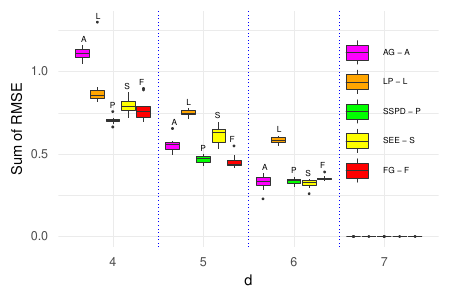}
\end{subfigure}
\begin{subfigure}{.5\textwidth}
\caption{Switching ripple: 5 objectives}
\includegraphics[width=\textwidth]{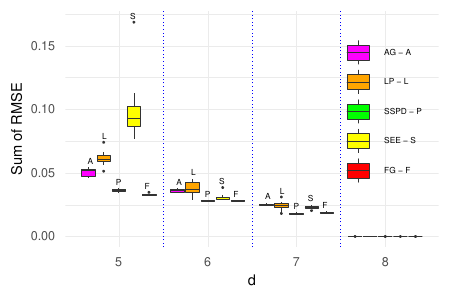}
\end{subfigure}%
\begin{subfigure}{.5\textwidth}
\caption{Switching ripple: 10 objectives}
\includegraphics[width=\textwidth]{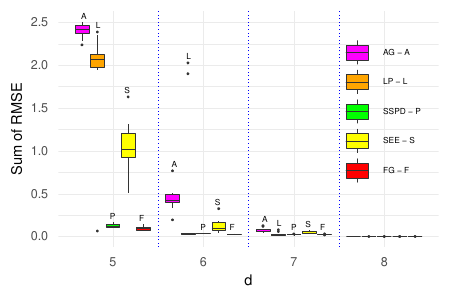}
\end{subfigure}
\caption{Comparison of the proposed approaches based on data sampled from the uniform distribution. The experiments were repeated $10$ times with the sample size $n=1000$ in each of them.}
\label{fig:boxplot_as_uniform}
\end{figure}

In what follows, AG stands for the average of gradients, MCH for the minimum norm element of the convex hull of the gradients, LP for the linear projection of gradients, SSPD for the sum of the SPD matrices, SEE for the stepwise estimation of eigenvectors, FG for Flury-Gautschi algorithm, and finally, Zahm for the method of \cite{Zahm20}. The results are illustrated as boxplots in Figure~\ref{fig:boxplot_as_normal}, for the standard normal, and in Figure~\ref{fig:boxplot_as_uniform}, for the uniform distribution, representing the sum of RMSEs \eqref{err:sum_root_mean_square_error} against the dimensionality $j$ of the dominating eigenspace $\hat{\W}_j$ defining a shared active subspace (different for each method). Note that, the vectors in $\hat{\W}_j$ are ordered by their importance. Because of space constraints, for all real-world problems, we have shown only the last $4$ dimensions. 

In all figures, one can easily observe that, as the number of dimensions increases, the error \eqref{err:sum_root_mean_square_error}  of all methods decreases, converging to zero for $\hat{\W}_d$, which contains all the eigenvectors, demonstrating the soundness of these methods. In the case of the normal distribution, for most problems and the dimensionality $j$, the method of Zahm performs the best, followed by the SPD-level methods.  Both for the standard normal and the uniform distribution, the SPD-level methods outperform the gradient-level ones and perform very similarly to each other, exhibiting a greater stability. From the gradient-level methods, the minimum-norm element of the convex hull exhibited the worst performance: although being positively correlated with all the gradients, having the minimum norm, it downplays the variability. Thus, we excluded this method from the graphs of all real-world optimization problems. It is noteworthy that the simple sum of SPD matrices provides similar to or better results than SSPD and FG, suggesting that it is a very useful instructive baseline.

\begin{figure}[htbp]
\centering
\includegraphics[width=10cm,height=4.7cm]{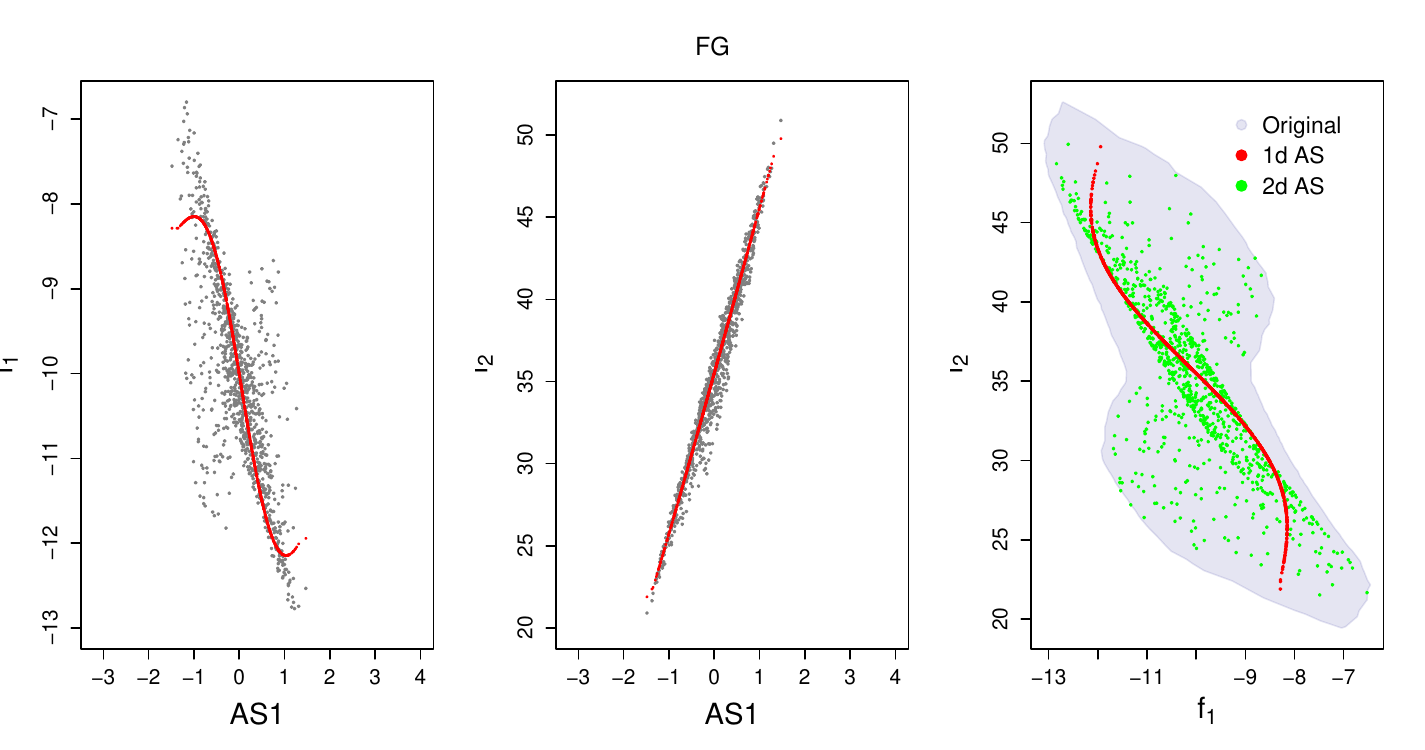}
\includegraphics[width=10cm,height=4.7cm]{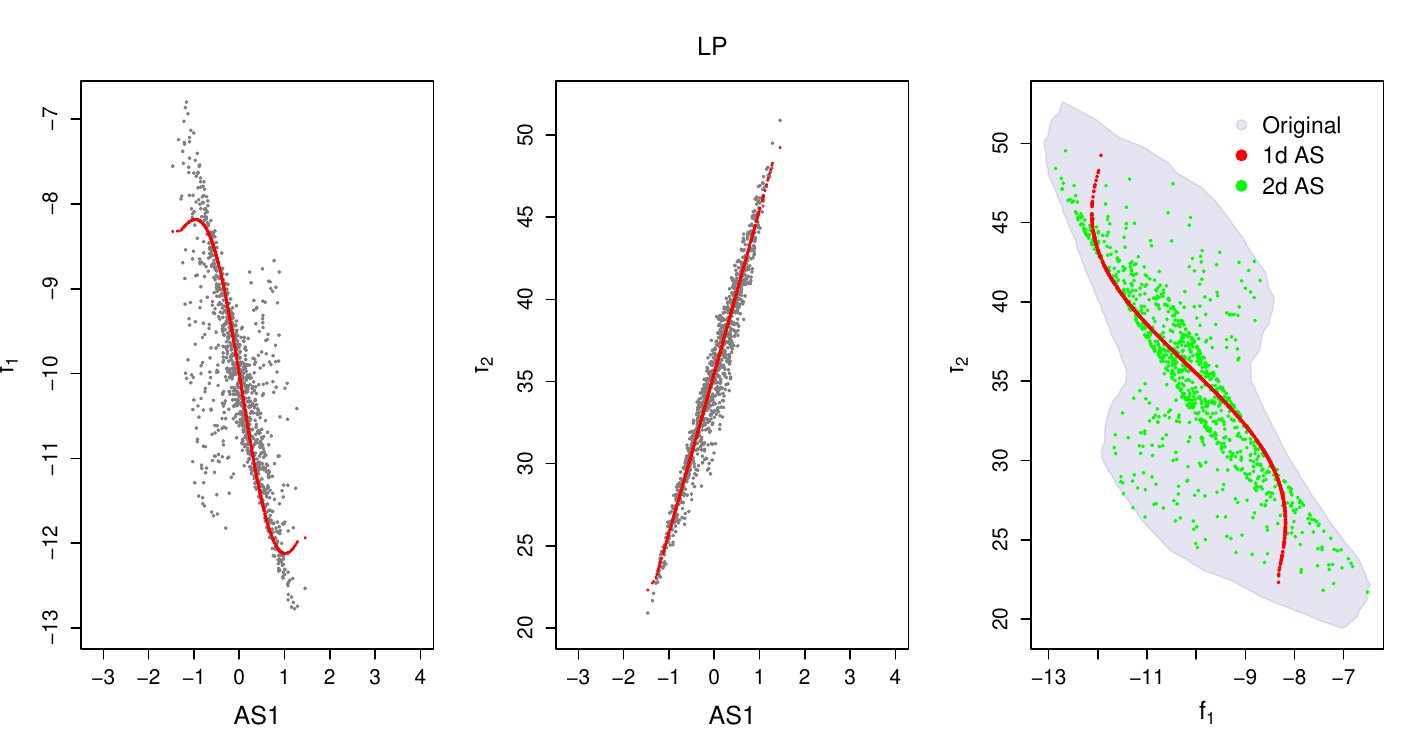}
\includegraphics[width=10cm,height=4.7cm]{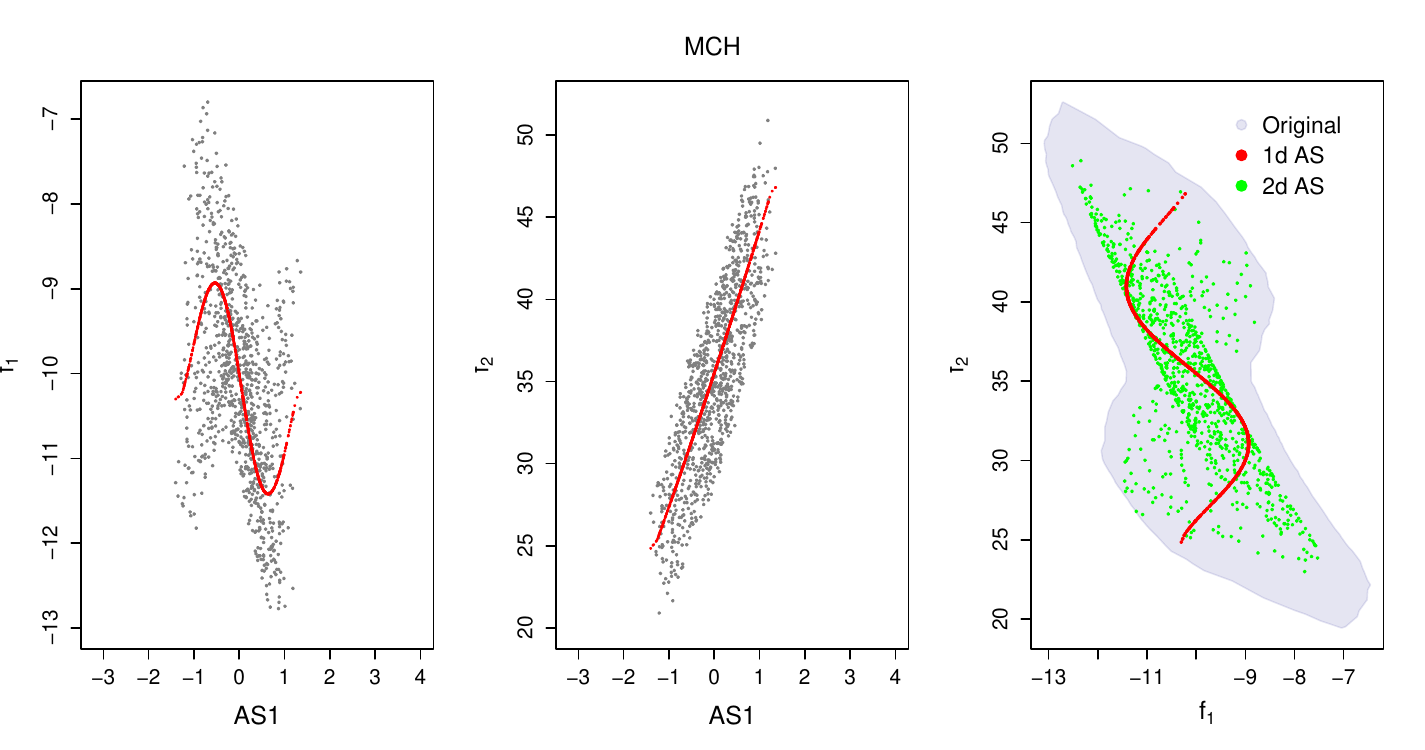}
\caption{Sufficient summary plots of a subset of methods for the synthetic problem applied to uniformly distributed data. AS1 stands for the active subspace of dimension $1$, and $f_1, f_2$ are the two objectives.}
\label{fig:summary_plot}
\end{figure}

Sufficient summary plot (see \cite{Constantine15}) is a useful tool to visualize the quality of the computed active subspace. Figure~\ref{fig:summary_plot} stands for such a plot for the synthetic problem where only a subset of methods has been shown; the rest is given in Figure~\ref{fig:summary_plot_rest} in Appendix~\ref{sec:summary_plots}. The first two plots in all panels illustrate the outputs of each function, respectively, on the given data $\D_n \in \R^{n \times d}$, marked as black ones, and  on the rotated data $\D_n\hat{\W}_1 \hat{\W}^T_1$, marked as red points, against the active subspace of dimension $1$, $\D_n\hat{\W}_1$. The third one reflects simultaneously the pairwise plot of the original outputs of the two component functions, and their outputs on $\D_n\hat{\W}_1 \hat{\W}^T_1$ (red points), and $\D_n\hat{\W}_2 \hat{\W}^T_2$ (green points). 
For this synthetic problem, the second objective $f_2$ demonstrates greater variability than $f_1$ wrt all input variables: this is captured by all methods, except AG and MCH, and is reflected in the form of a univariate trend in their corresponding (the middle) scatter plots. As the dimensionality $j$ is increased to $2$, the scatter plots in the rightmost graph for all methods, except for AG and MCH, represent well the original output space. This can also be verified from the summary given in Table~\ref{subtable:rmse_bc1}, and is inline with the boxplots of Figure~\ref{fig:boxplot_as_uniform}.
%Table~\ref{table:rmse_bc} shows that, overall, SEE performs better: objective wise it improves much the first one at the cost of degrading the second one compared to LP. Correspondingly, the scatter in the first plot of SEE is less than that of AG, which is not the case for the second one. 

Finally, as a remark, Table~\ref{table:rmse_bc} in the main body, and Table~\ref{tab:impact-norm} in Appendix~\ref{sec:effect-norm}, report the impact of the normalization of the function outputs when computing the gradients, on the RMSE performances of the methods: the performance improvement it provides, both in terms of the error and the variance, depends on the nature of the problem and the method considered, and the dimensionality of the active subspace. In particular, such a manipulation affects negatively the gradient-level methods (they are sensitive to the magnitudes of the gradients), whereas the SPD-level ones mostly benefit from it (they seek a common structure in the SPD matrices).

\begin{table}[h]
    \centering
   \scriptsize
        \begin{subtable}{0.45\linewidth} % Adjust the width as needed
        \centering
        \begin{tabular}{c|c|c|c|c|c|c}
            \multicolumn{4}{c}{j=1} & \multicolumn{3}{c}{j=2} \\
            \hline
            Method & $f_1$ & $f_2$ & sum & $f_1$ & $f_2$ & sum \\
            \hline
            AG & 0.969 & 0.994 & 1.963 & 0.609 & 0.156 & 0.765 \\
            MCH & 0.817 & 0.606 & 1.423 & 0.545 & 0.296 & 0.841 \\
            \textbf{LP}& 0.739 & 0.210 & 0.949 & \textbf{0.129} & \textbf{0.182} & \textbf{0.311} \\
            SSPD & 0.756 & 0.281 & 1.037 & 0.171 & 0.146 & 0.317 \\
            SEE & 0.758 & 0.259 & 1.017 & 0.245 & 0.083 & 0.328 \\
            \textbf{FG} &\textbf{0.746} & \textbf{0.188} & \textbf{0.934} & 0.166 &0.155 & 0.321 \\
            \hline
        \end{tabular}
        \caption{}
        \label{subtable:rmse_bc1}
    \end{subtable}
  \hspace{0.8cm}
    \begin{subtable}{0.45\linewidth} % Adjust the width as needed
        \centering
        \begin{tabular}{c|c|c|c|c|c|c}
            \multicolumn{4}{c}{j=1} & \multicolumn{3}{c}{j=2} \\
            \hline
            Method & $f_1$ & $f_2$ & sum & $f_1$ & $f_2$ & sum \\
            \hline
            AG & 0.774 & 0.135 & 0.909 & 0.373 & 0.102 & 0.475 \\
            MCH & 1.207 & 1.228 & 2.435 & 1.195 & 1.080 & 2.275 \\
            \textbf{LP} & \textbf{0.699} & \textbf{0.056} & \textbf{0.755} & 0.532 & 0.046 & 0.578 \\
            SSPD & 0.707 & 0.057 & 0.764 & 0.292 & 0.048 & 0.340 \\
            \textbf{SEE} & 0.741 & 0.256 & 0.997 & \textbf{0.246} & \textbf{0.081} & \textbf{0.327} \\
            FG & 0.706 & 0.056 & 0.762 & 0.317 & 0.050 & 0.367 \\
            \hline
        \end{tabular}
        \caption{}
        \label{subtable:rmse_bc2}
    \end{subtable}
 \caption{Synthetic problem: the objective-wise RMSEs, $f_1$, $f_2$, and the sum. In (b) the gradients are computed from the original outputs, whereas in (a) from the mean-variance normalized outputs. The best performing method in (each table separately) is shown in bold.}
    \label{table:rmse_bc}
\end{table}

%\begin{figure}[htbp]
%\centering
%\includegraphics[width=17cm, height=10cm]{penicilin_gradients.pdf}
%\caption{Penicillin vs carbon dioxide gradients.}
%\label{fig:penicillin_gradients}
%\end{figure}

\section{Conclusion and perspectives} \label{sec:conclusions}
We proposed several baseline gradient-based methods to compute a shared active subspace for multivariate vector-valued functions. These methods aim to reduce the vector-valued problem 
to the scalar-valued setting for which there exists an effective active subspace method, either at the level of the gradients or at the level of the SPD matrices computed from these gradients. They can be applied to any data regardless of the underlying distribution unlike the existing vector-valued approach which is based on the distributions satisfying the Poincar\'e-type inequalities.
The experiments show that regardless the dimensionality of the active subspace, the SPD-level methods perform better, both in terms of the error and the stability, than the gradient-level ones, and close to the vector-valued approach in the case of a normal distribution. The noteworthy fact is that, in most cases, the best shared active subspace can be computed by the simple sum of the SPD matrices.

In this work we did not give any preference to objective values, neither did we take into account any particularity or structure in the problem of interest.
However, in some applications one might favor certain regions in the image space of the function of interest, for which it is pertinent to consider a dedicated shared active subspace. One of the approaches to this problem may be based on the use of data depth tools which measure the centrality of a given point in a data cloud (more in Appendix~\ref{sec:depth}). Another perspective, inspired from \cite{Spagnol19}, would be to combine seamlessly multi-objective optimization and dimensionality reduction.

%a more flexible dimensionality reduction approach based on an optimization, and not (necessarily) variability driven, goal.

\newpage
\bibliographystyle{chicago}
\bibliography{asbib}

\begin{thebibliography}{}

\bibitem[\protect\citeauthoryear{Beaghen}{Beaghen}{1997}]{Beaghen97}
Beaghen, M. (1997).
\newblock {\em Canonical variate analysis and related methods with longitudinal
  data}.
\newblock Ph.\ D. thesis, Virginia Polytechnic Institute and State University.

\bibitem[\protect\citeauthoryear{Browne and McNicholas}{Browne and
  McNicholas}{2014}]{Browne14}
Browne, R.~P. and P.~D. McNicholas (2014).
\newblock Estimating common principal components in high dimensions.
\newblock {\em Advances in Data Analysis and Classification\/}~{\em 8},
  217--226.

\bibitem[\protect\citeauthoryear{Constantine and Gleich}{Constantine and
  Gleich}{2015}]{Constantine14b}
Constantine, P. and D.~Gleich (2015).
\newblock Computing active subspaces with monte carlo.

\bibitem[\protect\citeauthoryear{Constantine}{Constantine}{2015}]{Constantine15}
Constantine, P.~G. (2015).
\newblock {\em Active subspaces: Emerging ideas for dimension reduction in
  parameter studies}.
\newblock SIAM.

\bibitem[\protect\citeauthoryear{Constantine, Dow, and Wang}{Constantine
  et~al.}{2014}]{Constantine14}
Constantine, P.~G., E.~Dow, and Q.~Wang (2014).
\newblock Active subspace methods in theory and practice: applications to
  kriging surfaces.
\newblock {\em SIAM Journal on Scientific Computing\/}~{\em 36\/}(4),
  A1500--A1524.

\bibitem[\protect\citeauthoryear{Cook}{Cook}{1996}]{Cook96}
Cook, R.~D. (1996).
\newblock Graphics for regressions with a binary response.
\newblock {\em Journal of the American Statistical Association\/}~{\em
  91\/}(435), 983--992.

\bibitem[\protect\citeauthoryear{Da~Veiga, Gamboa, Iooss, and Prieur}{Da~Veiga
  et~al.}{2021}]{Da21}
Da~Veiga, S., F.~Gamboa, B.~Iooss, and C.~Prieur (2021).
\newblock {\em Basics and trends in sensitivity analysis: Theory and practice
  in R}.
\newblock SIAM.

\bibitem[\protect\citeauthoryear{Deb, Gupta, Daum, Branke, Mall, and
  Padmanabhan}{Deb et~al.}{2009}]{Deb09}
Deb, K., S.~Gupta, D.~Daum, J.~Branke, A.~K. Mall, and D.~Padmanabhan (2009).
\newblock Reliability-based optimization using evolutionary algorithms.
\newblock {\em IEEE transactions on evolutionary computation\/}~{\em 13\/}(5),
  1054--1074.

\bibitem[\protect\citeauthoryear{D{\'e}sid{\'e}ri}{D{\'e}sid{\'e}ri}{2012}]{Desideri12}
D{\'e}sid{\'e}ri, J.-A. (2012).
\newblock Multiple-gradient descent algorithm (mgda) for multiobjective
  optimization.
\newblock {\em Comptes Rendus Mathematique\/}~{\em 350\/}(5-6), 313--318.

\bibitem[\protect\citeauthoryear{Flury}{Flury}{1988}]{Flury88}
Flury, B. (1988).
\newblock {\em Common principal components \& related multivariate models}.
\newblock John Wiley \& Sons, Inc.

\bibitem[\protect\citeauthoryear{Flury}{Flury}{1987}]{Flury87}
Flury, B.~K. (1987).
\newblock Two generalizations of the common principal component model.
\newblock {\em Biometrika\/}~{\em 74\/}(1), 59--69.

\bibitem[\protect\citeauthoryear{Flury}{Flury}{1984}]{Flury1984}
Flury, B.~N. (1984).
\newblock Common principal components in k groups.
\newblock {\em Journal of the American Statistical Association\/}~{\em
  79\/}(388), 892--898.

\bibitem[\protect\citeauthoryear{Flury and Gautschi}{Flury and
  Gautschi}{1986}]{Flury86}
Flury, B.~N. and W.~Gautschi (1986).
\newblock An algorithm for simultaneous orthogonal transformation of several
  positive definite symmetric matrices to nearly diagonal form.
\newblock {\em SIAM Journal on Scientific and Statistical Computing\/}~{\em
  7\/}(1), 169--184.

\bibitem[\protect\citeauthoryear{Fukumizu and Leng}{Fukumizu and
  Leng}{2014}]{Fukumizu14}
Fukumizu, K. and C.~Leng (2014).
\newblock Gradient-based kernel dimension reduction for regression.
\newblock {\em Journal of the American Statistical Association\/}~{\em
  109\/}(505), 359--370.

\bibitem[\protect\citeauthoryear{Gilbert, Varadhan, and Gilbert}{Gilbert
  et~al.}{2009}]{Gilbert09}
Gilbert, P., R.~Varadhan, and M.~P. Gilbert (2009).
\newblock Package ‘numderiv’.
\newblock {\em differential equations\/}~{\em 3}, 203--267.

\bibitem[\protect\citeauthoryear{Guyon and Elisseeff}{Guyon and
  Elisseeff}{2003}]{Guyon03}
Guyon, I. and A.~Elisseeff (2003).
\newblock An introduction to variable and feature selection.
\newblock {\em Journal of machine learning research\/}~{\em 3\/}(Mar),
  1157--1182.

\bibitem[\protect\citeauthoryear{He, Tian, Wang, and Jin}{He
  et~al.}{2020}]{He20}
He, C., Y.~Tian, H.~Wang, and Y.~Jin (2020).
\newblock A repository of real-world datasets for data-driven evolutionary
  multiobjective optimization.
\newblock {\em Complex \& Intelligent Systems\/}~{\em 6}, 189--197.

\bibitem[\protect\citeauthoryear{Holodnak, Ipsen, and Smith}{Holodnak
  et~al.}{2018}]{Holodnak18}
Holodnak, J.~T., I.~C. Ipsen, and R.~C. Smith (2018).
\newblock A probabilistic subspace bound with application to active subspaces.
\newblock {\em SIAM Journal on Matrix Analysis and Applications\/}~{\em
  39\/}(3), 1208--1220.

\bibitem[\protect\citeauthoryear{Ji, Wang, Zahm, Marzouk, Yang, Ren, and
  Law}{Ji et~al.}{2018}]{Ji18}
Ji, W., J.~Wang, O.~Zahm, Y.~M. Marzouk, B.~Yang, Z.~Ren, and C.~K. Law (2018).
\newblock Shared low-dimensional subspaces for propagating kinetic uncertainty
  to multiple outputs.
\newblock {\em Combustion and Flame\/}~{\em 190}, 146--157.

\bibitem[\protect\citeauthoryear{Krzanowski}{Krzanowski}{1979}]{Krzanowski79}
Krzanowski, W. (1979).
\newblock Between-groups comparison of principal components.
\newblock {\em Journal of the american statistical association\/}~{\em
  74\/}(367), 703--707.

\bibitem[\protect\citeauthoryear{Lee}{Lee}{2019}]{Lee19}
Lee, M.~R. (2019).
\newblock Modified active subspaces using the average of gradients.
\newblock {\em SIAM/ASA Journal on Uncertainty Quantification\/}~{\em 7\/}(1),
  53--66.

\bibitem[\protect\citeauthoryear{Legendre, Oksanen, and ter Braak}{Legendre
  et~al.}{2011}]{Legendre11}
Legendre, P., J.~Oksanen, and C.~J. ter Braak (2011).
\newblock Testing the significance of canonical axes in redundancy analysis.
\newblock {\em Methods in Ecology and Evolution\/}~{\em 2\/}(3), 269--277.

\bibitem[\protect\citeauthoryear{Liang and Lai}{Liang and Lai}{2021}]{Liang21}
Liang, Q. and L.~Lai (2021).
\newblock Scalable bayesian optimization accelerates process optimization of
  penicillin production.
\newblock In {\em NeurIPS 2021 AI for Science Workshop}.

\bibitem[\protect\citeauthoryear{Mosler}{Mosler}{2013}]{Mosler13}
Mosler, K. (2013).
\newblock {\em Depth Statistics}, pp.\  17–34.
\newblock Springer Berlin Heidelberg.

\bibitem[\protect\citeauthoryear{Parr}{Parr}{2013}]{parr2013improvement}
Parr, J. (2013).
\newblock {\em Improvement criteria for constraint handling and multiobjective
  optimization}.
\newblock Ph.\ D. thesis, University of Southampton.

\bibitem[\protect\citeauthoryear{Parsons and Scott}{Parsons and
  Scott}{2004}]{Parsons04}
Parsons, M.~G. and R.~L. Scott (2004).
\newblock Formulation of multicriterion design optimization problems for
  solution with scalar numerical optimization methods.
\newblock {\em Journal of Ship Research\/}~{\em 48\/}(01), 61--76.

\bibitem[\protect\citeauthoryear{Pepler}{Pepler}{2014}]{Pepler14}
Pepler, P.~T. (2014).
\newblock {\em The identification and application of common principal
  components}.
\newblock Ph.\ D. thesis, Stellenbosch: Stellenbosch University.

\bibitem[\protect\citeauthoryear{Saltelli, Ratto, Andres, Campolongo, Cariboni,
  Gatelli, Saisana, and Tarantola}{Saltelli et~al.}{2008}]{Saltelli08}
Saltelli, A., M.~Ratto, T.~Andres, F.~Campolongo, J.~Cariboni, D.~Gatelli,
  M.~Saisana, and S.~Tarantola (2008).
\newblock {\em Global sensitivity analysis: the primer}.
\newblock John Wiley \& Sons.

\bibitem[\protect\citeauthoryear{Sekitani and Yamamoto}{Sekitani and
  Yamamoto}{1993}]{Sekitani93}
Sekitani, K. and Y.~Yamamoto (1993).
\newblock A recursive algorithm for finding the minimum norm point in a
  polytope and a pair of closest points in two polytopes.
\newblock {\em Mathematical Programming\/}~{\em 61}, 233--249.

\bibitem[\protect\citeauthoryear{Spagnol, Riche, and Veiga}{Spagnol
  et~al.}{2019}]{Spagnol19}
Spagnol, A., R.~L. Riche, and S.~D. Veiga (2019).
\newblock Global sensitivity analysis for optimization with variable selection.
\newblock {\em SIAM/ASA Journal on uncertainty quantification\/}~{\em 7\/}(2),
  417--443.

\bibitem[\protect\citeauthoryear{Tanabe and Ishibuchi}{Tanabe and
  Ishibuchi}{2020}]{Tanabe20}
Tanabe, R. and H.~Ishibuchi (2020).
\newblock An easy-to-use real-world multi-objective optimization problem suite.
\newblock {\em Applied Soft Computing\/}~{\em 89}, 106078.

\bibitem[\protect\citeauthoryear{Trendafilov and Gallo}{Trendafilov and
  Gallo}{2021}]{Trendafilov21}
Trendafilov, N. and M.~Gallo (2021).
\newblock {\em Multivariate data analysis on matrix manifolds}.
\newblock Springer.

\bibitem[\protect\citeauthoryear{Trendafilov}{Trendafilov}{2010}]{Trendafilov10}
Trendafilov, N.~T. (2010).
\newblock Stepwise estimation of common principal components.
\newblock {\em Computational Statistics \& Data Analysis\/}~{\em 54\/}(12),
  3446--3457.

\bibitem[\protect\citeauthoryear{Tukey}{Tukey}{1975}]{Tukey75}
Tukey, J.~W. (1975).
\newblock Mathematics and the picturing of data.
\newblock In {\em Proceedings of the International Congress of Mathematicians,
  Vancouver, 1975}, Volume~2, pp.\  523--531.

\bibitem[\protect\citeauthoryear{Wycoff, Binois, and Wild}{Wycoff
  et~al.}{2021}]{wycoff2021sequential}
Wycoff, N., M.~Binois, and S.~M. Wild (2021).
\newblock Sequential learning of active subspaces.
\newblock {\em Journal of Computational and Graphical Statistics\/}~{\em
  30\/}(4), 1224--1237.

\bibitem[\protect\citeauthoryear{Zahm, Constantine, Prieur, and Marzouk}{Zahm
  et~al.}{2020}]{Zahm20}
Zahm, O., P.~G. Constantine, C.~Prieur, and Y.~M. Marzouk (2020).
\newblock Gradient-based dimension reduction of multivariate vector-valued
  functions.
\newblock {\em SIAM Journal on Scientific Computing\/}~{\em 42\/}(1),
  A534--A558.

\bibitem[\protect\citeauthoryear{Zhang, He, Ye, Xu, and Pan}{Zhang
  et~al.}{2019}]{Zhang19}
Zhang, Z., C.~He, J.~Ye, J.~Xu, and L.~Pan (2019).
\newblock Switching ripple suppressor design of the grid-connected inverters: A
  perspective of many-objective optimization with constraints handling.
\newblock {\em Swarm and evolutionary computation\/}~{\em 44}, 293--303.

\end{thebibliography}

\appendix

\newpage
\section{Summary plots for the synthetic problem}
\label{sec:summary_plots}
\begin{figure}[htbp]
\centering
\includegraphics[width=10.5cm,height=4.6cm]{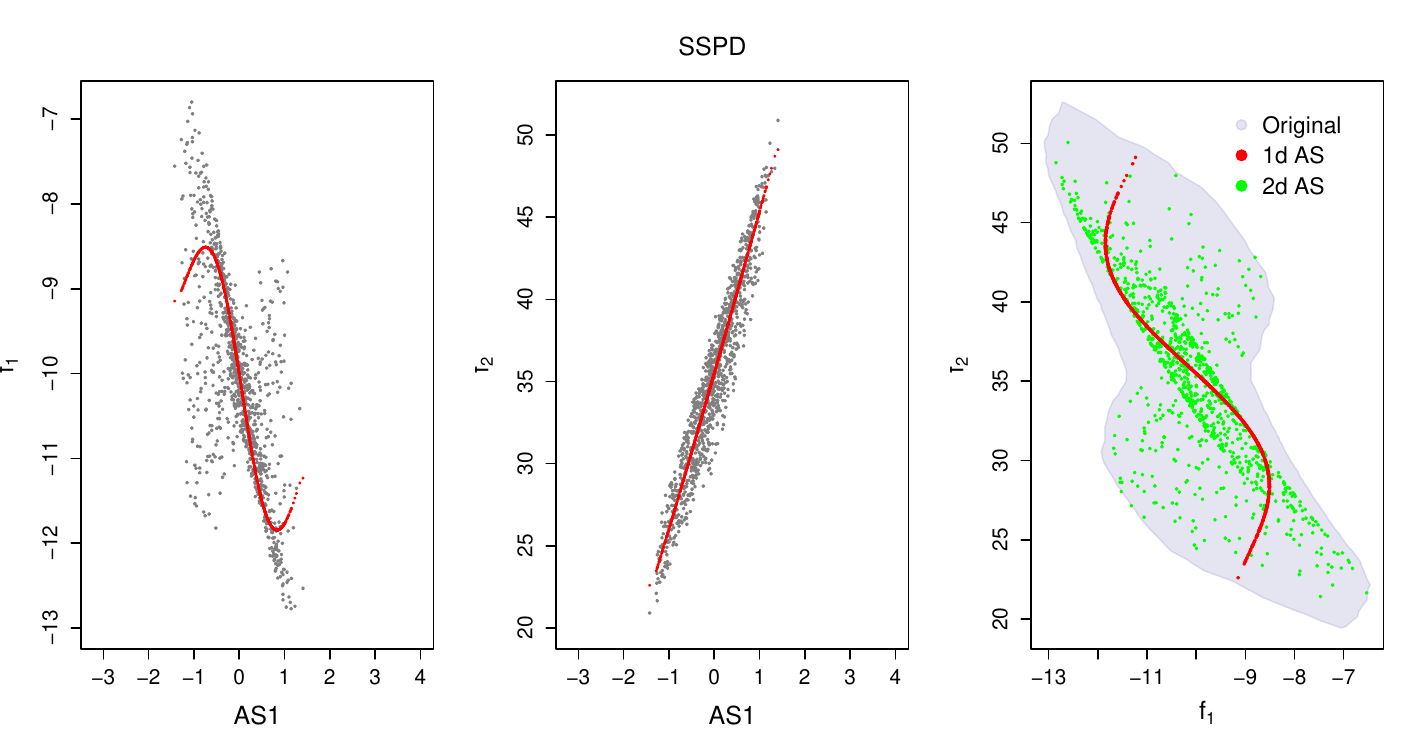}
\includegraphics[width=10.5cm,height=4.6cm]{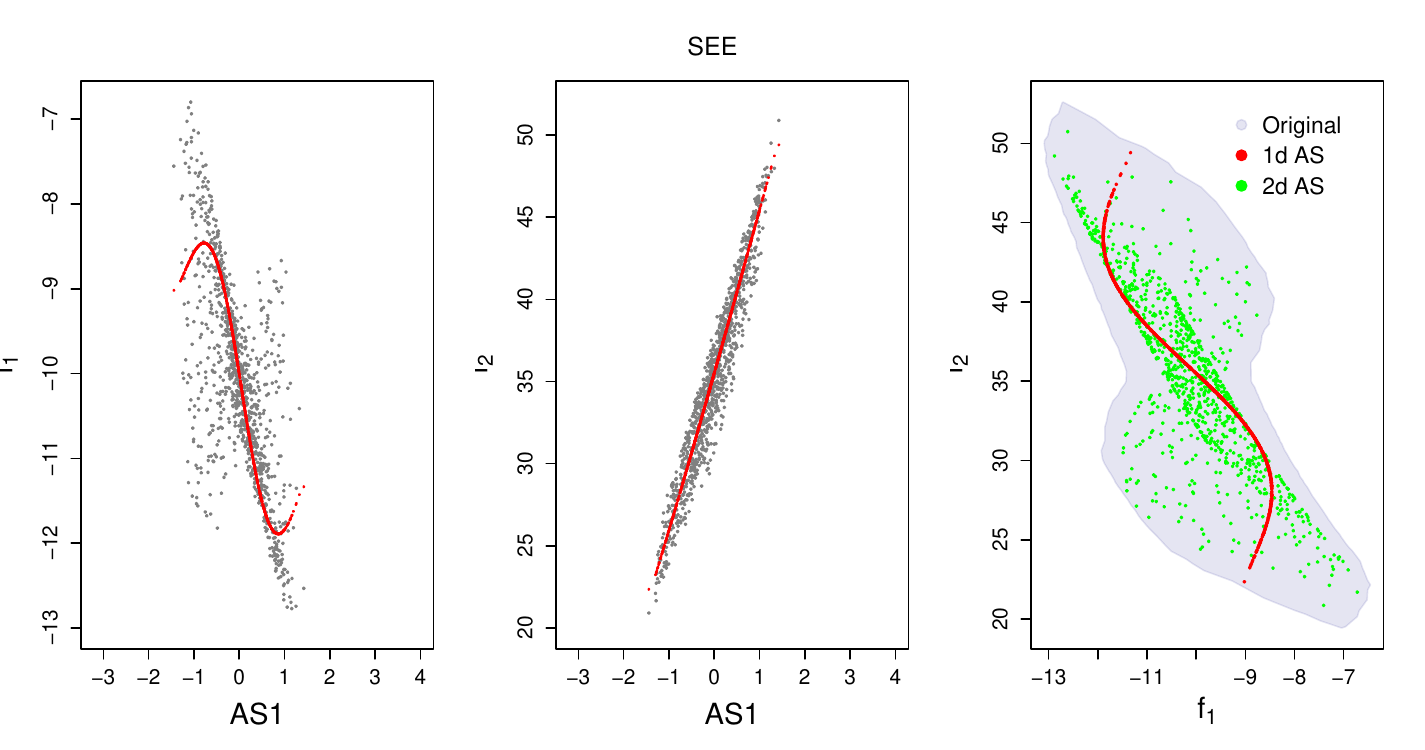}
\includegraphics[width=10.5cm,height=4.6cm]{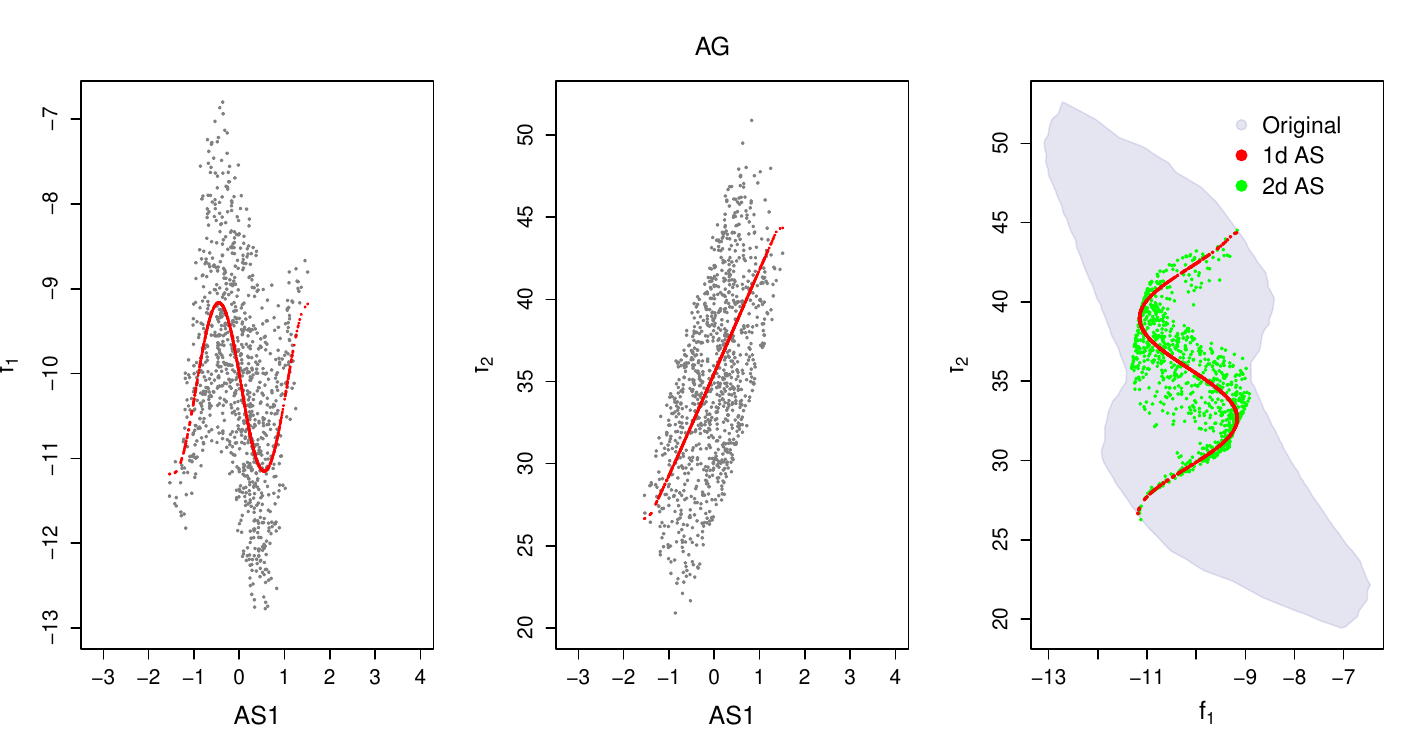}
\caption{Summary plots of SSPD, SEE, and AG for the synthetic problem applied to uniformly distributed data. AS1 stands for the active subspace of dimension $1$, and $f_1, f_2$ are the two objectives.}
\label{fig:summary_plot_rest}
\end{figure}

\newpage
\section{Impact of normalization on the RMSE performance}
\label{sec:effect-norm}
\begin{table}[htbp]
  \caption{Impact of the mean-variance normalization for a subset of problems. The values are the averages of the sum of RMSEs over $10$ runs with the sample size $n=1000$ in each of them. $j$ stands for the dimensionality of the leading eigenspace: here are given the performances with respect to the two penultimate eigenspaces. The better performances are shown in bold.}
  \vspace{-0.5cm}
\centering
\scriptsize
  \begin{subtable}{0.4\textwidth}
    \centering
    \caption{Normalized car side-impact}
    \begin{tabular}{c c c}
      Method & j=5 & j=6 \\
      \hline
      AG & 3.688 & 2.920 \\
      LP & 4.712 & 3.872 \\
      SSPD & \textbf{1.266} & \textbf{0.660} \\
      SEE & \textbf{1.680} & \textbf{0.770} \\
      FG & \textbf{1.312} & \textbf{0.764} \\
      \hline
    \end{tabular}

    \label{tab:cartable1}
  \end{subtable}
  \hspace{-1cm}
  \begin{subtable}{0.4\textwidth}
    \centering
    \caption{Original car side-impact}
    \begin{tabular}{c c c}
      Method & j=5 & j=6 \\
      \hline
      AG & \textbf{3.128} & \textbf{2.262} \\
      LP & \textbf{3.260} & \textbf{2.329} \\
      SSPD & 1.974 & 1.549 \\
      SEE & 1.682 & 0.834 \\
      FG & 2.012 & 1.574 \\
      \hline
    \end{tabular}
    \label{tab:cartable2}
  \end{subtable}
 %%%%%%%%%%%%%%%%%%%%%%%%%%%%%%%%%%%%%%%%%
  \begin{subtable}{0.45\textwidth}
    \centering
\caption{Normalized penicillin production}
    \begin{tabular}{c c c}
      Method & j=5 & j=6 \\
      \hline
      AG & \textbf{0.556} & 0.327 \\
      LP & \textbf{0.748} & 0.583 \\
      SSPD & \textbf{0.467} & 0.336 \\
      SEE & \textbf{0.615} & 0.322 \\
      FG & \textbf{0.452} & 0.353 \\
      \hline
    \end{tabular}
    \label{tab:pentable1}
  \end{subtable}
  \hspace{-2cm}
  \begin{subtable}{0.45\textwidth}
    \centering
        \caption{Original penicillin production}
    \begin{tabular}{c c c}
      Method & j=5 & j=6 \\
      \hline
      AG & 0.779 & \textbf{0.321} \\
      LP & 0.777 & \textbf{0.220} \\
      SSPD & 0.764 & \textbf{0.211} \\
      SEE & 0.655 & \textbf{0.280} \\
      FG & 0.717 & \textbf{0.193} \\
      \hline
    \end{tabular}
    \label{tab:pentable2}
   \end{subtable}
%%%%%%%%%%%%%%%%%%%%%%%%%%%%%%%%%%%%%%%%
   \begin{subtable}{0.45\textwidth}
       \caption{Normalized switching ripple}
    \centering
    \begin{tabular}{c c c}
      Method & j=6 & j=7 \\
      \hline
      AG & 0.036 & \textbf{0.025} \\
      LP & 0.038 & \textbf{0.024} \\
      SSPD & \textbf{0.028} & \textbf{0.018} \\
      SEE & 0.031 & 0.023 \\
      FG & \textbf{0.028} & \textbf{0.019} \\
      \hline
    \end{tabular}
    \label{tab:srippletable1}
  \end{subtable}
  \hspace{-2cm}
  \begin{subtable}{0.45\textwidth}
    \centering
        \caption{Original switching ripple}
    \begin{tabular}{c c c}
      Method & j=6 & j=7 \\
      \hline
      AG & \textbf{0.032} & 0.029 \\
      LP & \textbf{0.034} & 0.029 \\
      SSPD & 0.031 & 0.028 \\
      SEE & 0.031 & \textbf{0.021} \\
      FG & 0.031 & 0.028 \\
      \hline
    \end{tabular}
    \label{tab:srippletable2}
  \end{subtable}
  \label{tab:impact-norm}
\end{table}

\section{Local active subspace based on data depth} \label{sec:depth}

Given a data $\D_n$, a data depth $D(\x | \D_n): \R^d \to [0,1]$ measures how central (or deep) a point is in this data cloud \cite{Tukey75}. It attains its maximum value for the deepest point in the cloud (for instance, the center of a symmetric distribution) and tends to zero for outlying points, i.e., it orders data by their degree of centrality. It possesses useful properties such as translation invariance $D(\x + \mathbf{b} | \D_n + \mathbf{b}) = D(\x | \D_n)$ for any $\mathbf{b} \in \R^d$, linear invariance  $D(\mathbf{A} \x | \mathbf{A} \D_n) = D(\x | \D_n)$ for every nonsingular $d \times d$ matrix $\mathbf{A}$ and so on \cite{Mosler13}. %Depth functions have been applied to a variety of problems, such as risk measure estimation \cite{Armaut22}, classification \cite{Pokotylo19b}, multi-objective optimization by \cite{Horiguchi22}. 
%Being a scalar-valued function, it is an attractive option to be considered also in the active subspace computation. 

\begin{figure}[htbp]
\centering
\includegraphics[width=\textwidth]{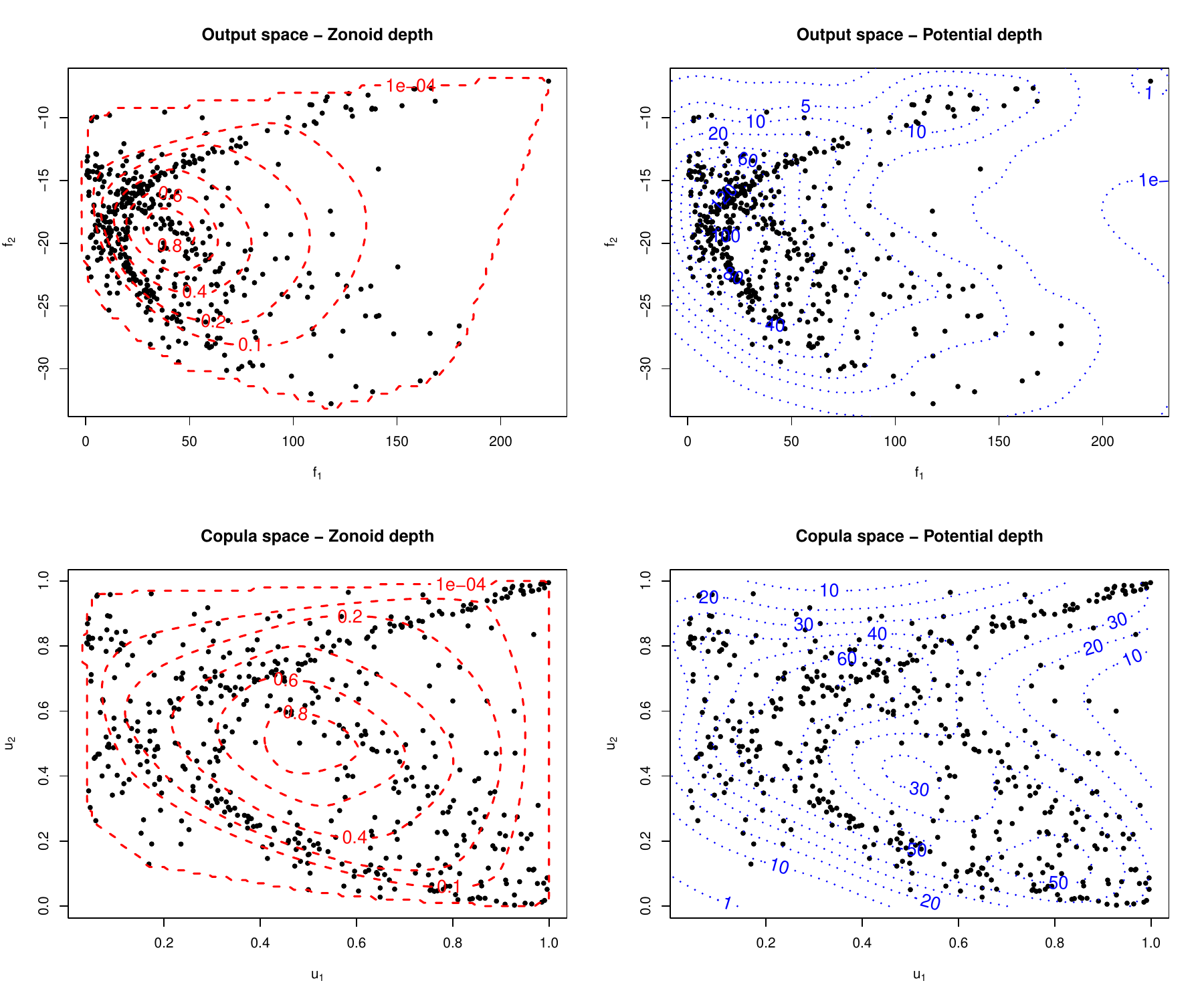}
\caption{Comparison of the zonoid \cite{Mosler13} and potential depths function on a test problem, in the original output space and in copula (rank) space.}
\label{fig:depthcomp}
\end{figure}

From the perspective of this work, we observe that depth functions vary the most on the areas where the sample points are the denser, and vary much less in the sparser parts: this is in opposition with the objective-wise gradients, since large gradients of the depths may correspond to the areas where the functions vary less (and hence cluster), unless the function is multimodal. One can alleviate this problem by applying a depth function in the copula or rank space which amounts to applying the inverse cumulative distribution function of every objective, making the depth value invariant to any increasing monotonous transformation of the objectives. As a result it puts more weight on the parts where the functions change the most as illustrated in Figure~\ref{fig:depthcomp}, and thus it may be used as a weight function to favor the given parts of the image space of the function of interest.

\end{document}